\documentclass[12pt]{article}      
\usepackage{graphicx}

\usepackage{amsmath,amsfonts,amssymb}

\usepackage{graphicx}
\usepackage{dcolumn}
\usepackage{bm}

\begin{document}


\title{Exact fixation probabilities for the Birth-Death and Death-Birth frequency-dependent Moran processes on the star graph}
\author{Evandro P. de Souza$^{1}$ \\Armando G. M. Neves$^{2}$
	\\
	\normalsize{$^{1}$Departamento de Matem\'atica, Universidade Federal de Minas Gerais}
	\\ 	\normalsize{evandrpds@gmail.com}\\
	\normalsize{$^{2}$Departamento de Matem\'atica, Universidade Federal de Minas Gerais}	
	\\	\normalsize{aneves@mat.ufmg.br}
}

\date{\today}
\maketitle
\begin{abstract}
Broom and Rycht\'{a}\v{r} [Proc. R. Soc. A (2008) 464, 2609--2627] found an exact	solution for the fixation probabilities of the Moran process for a structured population, in	which the interaction structure among individuals is given by the so-called star graph, i.e.	one central vertex and $n$ leaves, the leaves connecting only to the center. We generalize on their solution by allowing individuals'fitnesses to depend on the population frequency, and also by allowing a possible change in the order of reproduction and death draws. In their cited paper, Broom and Rycht\'{a}\v{r} considered the birth-death (BD) process, in which at each time step an individual is first drawn for reproduction and then an individual is selected for death. In the death-birth (DB) process, the order of the draws is reversed. It may be seen that the order of the draws makes a big difference in the fixation probabilities. Our solution method applies to both	the BD and the DB cases. As expected, the exact formulae for the fixation probabilities are complicated. We will also illustrate them with some examples and provide results on the	asymptotic behavior of the fixation probabilities when the number $n$ of leaves in the graph tends to infinity.
\end{abstract}

\section{Introduction} \label{secintro}
The Moran process \cite{moran} is a \textit{stochastic} model for the evolution of a \textit{finite} population of several types of individuals with different fitnesses, asexual reproduction and \textit{no mutations}. Another model for the same situation is the Wright-Fisher process \cite{fisher, wright}. Mathematically speaking, both models are discrete-time Markov chains \cite{allen} with a finite set of states. Due to the no-mutations hypothesis, the states in which all individuals are of the same type are absorbing. As all the other states are transient, with probability 1 and after sufficient time, one of the absorbing states will be reached \cite{allen}. This is the phenomenon of \textit{fixation} and we say that the only type of individuals present in the final state is fixated in the population. One important problem is calculating as a function of the initial state of the population, i.e. how many individuals of each type is initially present, the probability of attaining each of the absorbing states.

One important difference between the Moran and Wright-Fisher processes is that in the simplest cases the fixation probabilities may be exactly calculated in the former, whereas in the latter we must use approximations \cite{ewens}. By simplest cases, we mean that the number of types of individuals is only 2 and that the population is \textit{fully mixed}. This exact solution, see eq. (\ref{exactpii}), is known since the original work by Moran \cite{moran}, and extends also to the case in which the fitnesses of the individuals depend on their population frequencies \cite{nowaknature, taylor}, i.e. Evolutionary Game Theory. If the number of types of individuals in the population is three or more \cite{wang2007evolutionary, moran3}, we can no longer calculate exactly the fixation probabilities, but we can find useful upper and lower bounds for them \cite{moran3}. As we are interested here in exact fixation probabilities, from now on we will consider only cases in which there are only individuals of two types in the population.

Most models in Mathematical Biology, e.g. the Lotka-Volterra predator-prey model, the SIR model for epidemics, the simpler versions of both Wright-Fisher and Moran models, and many others suppose that populations are \textit{fully mixed} or, in different words, have \textit{no structure}. This hypothesis means that all individuals in the population can interact with equal probability with any other individual. In real populations, on the contrary, there exist in general small communities and individuals with more or less contacts outside their community. In recent times, research focused on the statistical properties of networks of individuals in real populations \cite{newman}, and which may be the effects on the mathematical models of substituting the simple full-mixing hypothesis for realistic networks \cite{pastorvespignani1, pastorvespignani2}.

The Moran process in a structured population was introduced in \cite{lieberman}. Population structure is modeled by a \textit{directed} and \textit{weighted} graph. Several different classes of graphs were described in the above reference, in which edges may be directional or not, and weighted. Authors gave without proof asymptotic expressions (in the infinite population limit) for the fixation probability of a randomly placed mutant individual for star and super-star graphs in the case of frequency-independent fitnesses. In particular, these asymptotic expressions imply that both star and super-star are  \textit {amplifiers} of selection, i.e. one individual fitter (respectively less fit) than the rest of the population will fixate with larger (resp. smaller) probability than in an unstructured population. Many papers followed the introduction of the subject and reviews are available \cite{szabo, rocacuestasanchez, shakarianroosjohnson, allennowak}.

Broom and Rycht\'{a}\v{r} \cite{broomrychtar} developed further the theory, showing that, in general, fixation probabilities for the Moran process on a graph may be calculated by solving a huge system of $2^N$ linear equations, where $N$ is the population size, i.e. the number of nodes in the graph. But they showed that for symmetric graphs the number of equations to be solved may be much smaller. Approximate calculations of fixation probabilities in general graphs may use Monte Carlo simulations  \cite{barbosa} or other algorithms \cite{shakarianroosmoores, hindersin}.

\begin{figure}
	\begin{center}
	\includegraphics[width=0.4 \textwidth]{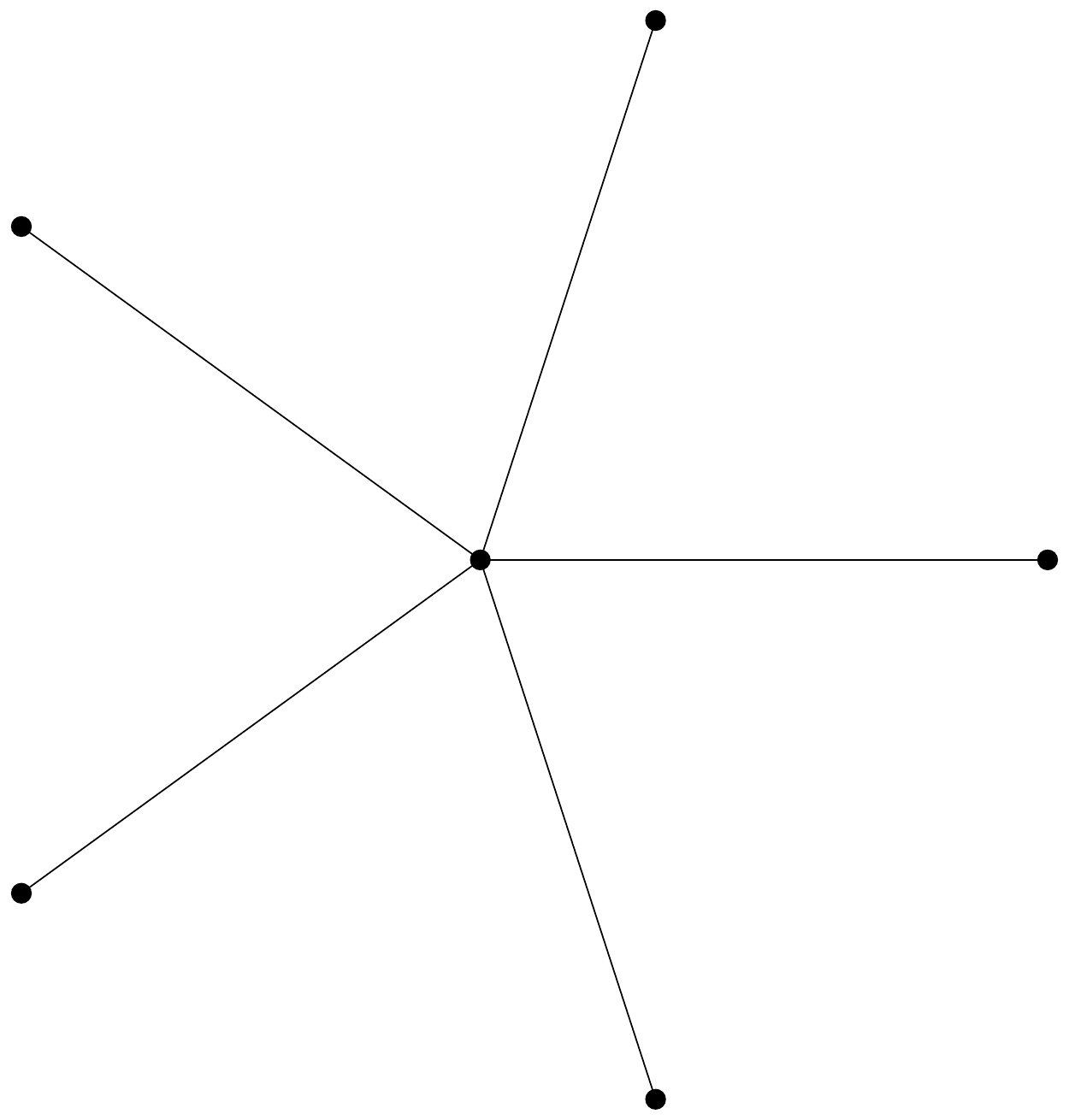}
	\caption{\label{figstar} The star graph with $n=5$ leaves.}
\end{center}
\end{figure}

One very simple symmetric graph in which the fixation probabilities were explicitly calculated  \cite{broomrychtar} is the \textit{star graph}. The star is a graph with $N=n+1$ vertices: the \textit{center} and $n$ leaves. The center is linked to all leaves and the leaves connect only to the center, see Fig. \ref{figstar}. All edges are bidirectional and all weights are equal.
The asymptotic formula in \cite{lieberman}, with a correction given by Chalub \cite{chalubstar}, follows from the exact solution of Broom and Rycht\'{a}\v{r}. A different derivation of the exact solution, using martingales, is given by Monk, Green and Paulin \cite{monk}. 

In this paper we will generalize on the exact result of \cite{broomrychtar, monk} for the fixation probabilities of the star graph. They were derived on the hypotheses of frequency-independent fitnesses and birth-death (BD) updating. Our derivation allows the fitnesses to depend on the population frequencies of the individuals and also the death-birth (DB) updating. We also provide explicit formulae for the fixation probabilities in the star graph for any initial configuration of A and B individuals, allowing us to study their asymptotic limits when the total population tends to infinity. 
	
In Sect. \ref{secMoran} we will introduce the Moran process for frequency-dependent fitnesses, both for structured and unstructured populations. For the former, we will define the BD and DB updating rules. In Sect. \ref{secExact} we will derive the exact expressions for the fixation probabilities in the star graph. In Sect. \ref{secAsymp} we derive, separately for each update rule, asymptotic expressions for the fixation probabilities in the limit of infinite populations. Some conclusions are drawn in Sect. \ref{secconc}.

\section{The Moran process}\label{secMoran}

In order to explain the Moran process on a structured population, we describe at first the standard, or \textit{unstructured}, Moran process. Consider a population with fixed size of $N$ individuals of two types A and B. Suppose that time is discrete and at each time step we draw random individuals, one for reproduction and one for death. The death lottery is uniform, but the reproduction lottery -- defined precisely below -- is performed in such a way that fitter individuals reproduce more frequently. The offspring of the reproducing individual is a single individual having the same type A or B as its parent. This is known as the \textit{no mutations} hypothesis. This offspring replaces the dead individual and population size $N$ thus remains constant. Unstructured means here that both lotteries are realized among all individuals in the population.

Let $f_i$ and $g_i$ be the fitnesses respectively of type A and type B individuals when the number of A individuals in the population is $i\in \{0, 1, \dots, N\}$. In the Evolutionary Game Theory context, these fitnesses are calculated in terms of a pay-off matrix \cite{nowakbook} and generally depend on $i$. In such a case, we say that fitnesses are frequency-dependent. In many cases we may consider that $f_i$ and $g_i$ do not depend on $i$ and we say that fitnesses are frequency-independent. The \textit{relative fitness} of A individuals is defined as
\begin{equation}  \label{defri}
r_i \,=\,\ \frac{f_i}{g_i}\;.
\end{equation}

The precise definition of the reproduction lottery is that the probabilities of drawing an A or a B for reproduction are respectively 
\begin{equation}  \label{replott}
\frac{i f_i}{i f_i+ (N-i)g_i}\;\;\textrm{and}\;\;\; \frac{(N-i)g_i}{i f_i+ (N-i)g_i}\;,
\end{equation}
i.e. proportional to the type's fitness.

It can be shown \cite{ewens,nowakbook} that the fixation probability of A individuals when their initial number is $i$ is exactly given by
\begin{equation}  \label{exactpii}
\pi_i= \frac{1+\sum_{j=1}^{i-1}\prod_{k=1}^{j} r_k^{-1}}{1+ \sum_{j=1}^{N-1}\prod_{k=1}^{j} r_k^{-1}}\;.
\end{equation}

We will now describe a particular case of the Moran process for a structured population. Let $\cal G$ be a graph with $N$ vertices. We suppose that there is exactly one individual at each vertex of $\cal G$. We interpret an edge linking vertices $a$ and $b$ as meaning that an offspring of the individual at $a$ can occupy the vertex $b$, and vice-versa. We say that $a$ and $b$ are neighbors in $\cal G$ if there is an edge between them. For a more general situation, in which the graph $\cal G$ describing the structure of the population is directed and weighted, and update rules other than BD and DB considered here, see e.g. \cite{allennowak}. For simplicity, we restrict our description to the case in which $\cal G$ is not directed and all edges have the same weight. The Moran process on $\cal G$ with BD updating is similar to the unstructured Moran process, but at each time step we first draw an individual for reproduction with probabilities given by (\ref{replott}) and then we draw an individual for death uniformly \textit{only among the neighbor vertices} in $\cal G$ of the reproducing individual. The offspring of the reproducing individual has the same type as its parent, \textit{no mutation} again, and occupies the vertex of the dead individual. In the DB updating, the order of the lotteries is reversed: we first draw with uniform probability an individual for dying and then, only among its neighbors in $\cal G$, we draw with probabilities proportional to fitness an individual to reproduce and have its offspring substitute the one that died. We will see soon that the order of the draws matters. In any case, the graph $\cal{G}$ provides a structure for the population, in which an individual does not necessarily interact with all other individuals. The standard Moran process is recovered if $\cal{G}$ is the complete graph on $N$ vertices with all edges having the same weight.

\section{Exact fixation probabilities in the star graph}\label{secExact} 
Let $n$ be the number of leaves in the star graph. A \textit{configuration} is described by the type A or B of the individual occupying the center and by the number $i \in \{0, 1, \dots, n\}$ of A individuals occupying the leaves. Let $(0,i)$ denote the configuration in which there is a B at the center and $i$ A individuals at the leaves. Accordingly, $(1,i)$ denotes the configuration in which there is an A at the center and $i$ As at the leaves. 

\begin{figure}
	\begin{center}
		\includegraphics[width=0.85 \textwidth]{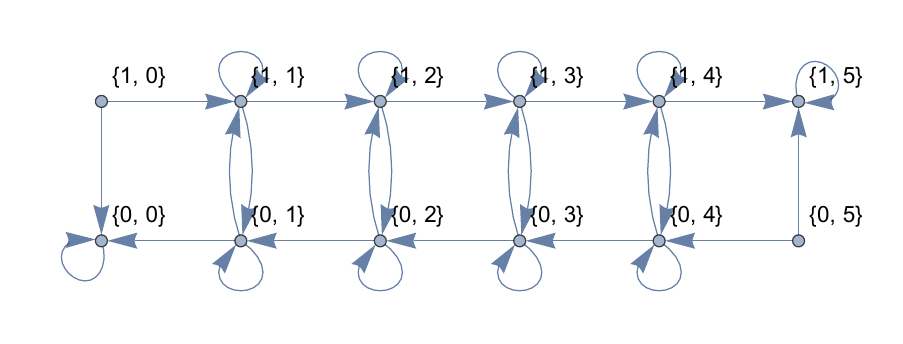}
		\caption{\label{startransitions} The possible transitions between configurations of the star graph with $n=5$ leaves.}
	\end{center}
\end{figure}

Fig. \ref{startransitions} illustrates, taking the case $n=5$ as an example, the possible configuration transitions in one time step, i.e. one reproduction and one death lottery, for the Moran process in the star graph for either BD or DB updates.
According to the figure, the lower configurations $(0,i)$ may in general move to the left neighbor $(0,i-1)$, upwards to $(1,i)$, or remain fixed. The \textit{minus} transition $(0,i) \rightarrow (0,i-1)$ happens when one draws the center vertex occupied by a B individual for reproduction and a leaf vertex occupied by an A for death. The probability of that transition is denoted $t_{i,-}$. Similarly, the \textit{up} transition $(0,i) \rightarrow (1,i)$ happens when one draws one of the $i$ leaves occupied by an A individual for reproduction and the center occupied by a B for death. The corresponding transition probability is denoted $t_{i,u}$. The other possible non-trivial transitions are the \textit{plus} transition $(1,i) \rightarrow (1,i+1)$, with probability $t_{i,+}$, and the down transition $(1,i) \rightarrow (0,i)$, with probability $t_{i,d}$. All the transition probabilities introduced above may be easily calculated according to the definitions of the process.

As an example, for the BD updating, $t_{i,+}$ is the probability of drawing for reproduction the center vertex occupied by an A times the probability of drawing any of the $n-i$ leaves occupied by a B  individual for death. The first of these, taking into account (\ref{replott}) and that the total number of A individuals in the population is $i+1$, is $r_{i+1}/[(i+1)r_{i+1}+n-i]$. The second probability, taking into account that the death lottery is uniform, is simply $(n-i)/n$. The complete set of non-trivial transition probabilities is given below. In deducing these formulae we also remind the reader that in the BD case the death probability for the center is 1 if a leaf is drawn for reproduction. In the DB case, the reproduction probability of the center is 1 if a leaf is drawn for death. The result is:

\textbf{BD case}:
\begin{align}
t_{i,u} &= \frac{ir_i}{ir_i+ n-i+1}  &t_{i,d} &= \frac{n-i}{(i+1)r_{i+1}+n-i} \nonumber \\
t_{i,-} &= \frac{1}{ir_i+n-i+1}\,\frac{i}{n} &t_{i,+} &= \frac{r_{i+1}}{(i+1)r_{i+1}+n-i}\,\frac{n-i}{n} \; \label{BDprobs}
\end{align}
\textbf{DB case}:
\begin{align}
t_{i,u} &= \frac{1}{n+1}\, \frac{ir_i}{ir_i+n-i} & t_{i,d} &= \frac{1}{n+1}\, \frac{n-i}{ir_i+n-i} \nonumber\\
t_{i,-} &= \frac{i}{n+1} &t_{i,+} &= \frac{n-i}{n+1} \;\label{DBprobs}
\end{align}

Let $P_i^0$ be the A fixation probability with initial condition $(0,i)$. 
Similarly, $P_i^1$ will denote the A fixation probability for initial condition $(1,i)$. Due to the no-mutation hypothesis, configurations $(0,0)$ and $(1,n)$, in which one single type is present, are absorbing. We have thus boundary conditions 
\begin{equation}
\label{bc}
P_0^0=0 \;\; \textrm{and} \;\; P_n^1=1\;.
\end{equation}

The equations for calculating the fixation probabilities in both BD and DB are
\begin{eqnarray}
P_i^0 &=& t_{i,u} P_i^1 + t_{i,-} P_{i-1}^0+ (1-t_{i,u}-t_{i,-})P_i^0 \nonumber\\
P_i^1 &=& t_{i,d} P_i^0 + t_{i,+} P_{i+1}^1+ (1-t_{i,d}-t_{i,+})P_i^1 \nonumber \;,
\end{eqnarray}
where $i$ runs between 1 and $n$ for the first line and from 0 to $n-1$ in the second. In order to find the fixation probabilities, we have thus to solve the above system of $2n$ equations, taking into account the boundary conditions (\ref{bc}).

The above equations can be rewritten as 
\begin{eqnarray}
P^0_i&=&\beta_i P^0_{i-1}+(1-\beta_i)P^1_i \label{eqp0i} \\
P^1_i&=&\alpha_i P^1_{i+1}+(1-\alpha_i)P^0_i \label{eqp1i}\;,
\end{eqnarray}
with 
\begin{equation}
\label{defalphabeta}
\beta_i= \frac{t_{i,-}}{t_{i,u}+t_{i,-}} \;\;\; \alpha_i= \frac{t_{i,+}}{t_{i,d}+t_{i,+}} \;.
\end{equation}

\begin{figure}
	\begin{center}
\includegraphics[width= \textwidth]{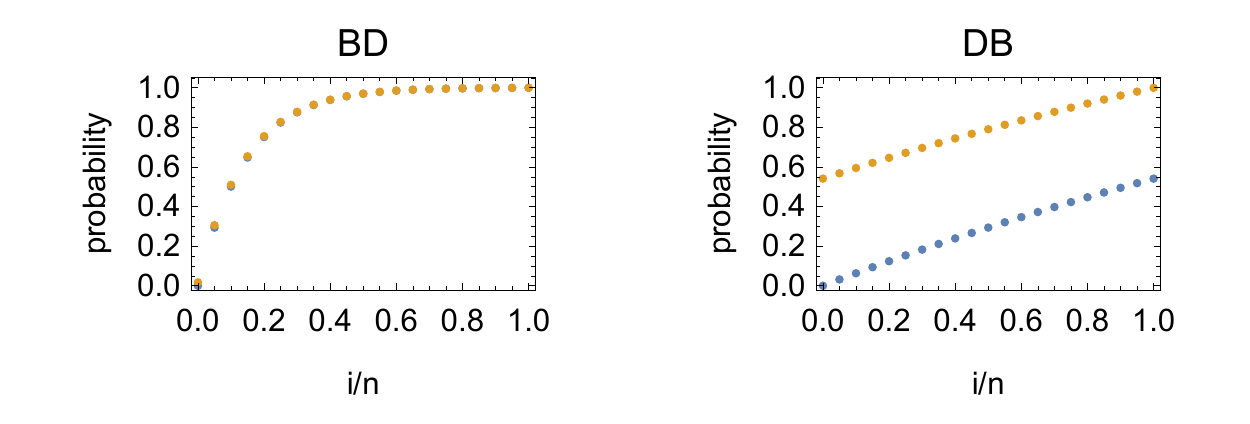}
\caption{\label{figindep} Left panel: plots of the fixation probabilities in the star graph with $n=20$ leaves, BD updating and frequency-independent relative fitness $r=1.2$. For the same value of $i$ the difference between $P^0_i$ and $P^1_i$ is so small that it is almost invisible. Right panel: the same for DB updating. In this case, the blue dots are the $P^0_i$, noticeably smaller than the $P^1_i$, represented by orange dots.}
	\end{center}
\end{figure}

\begin{figure}
	\begin{center}
		\includegraphics[width=\textwidth]{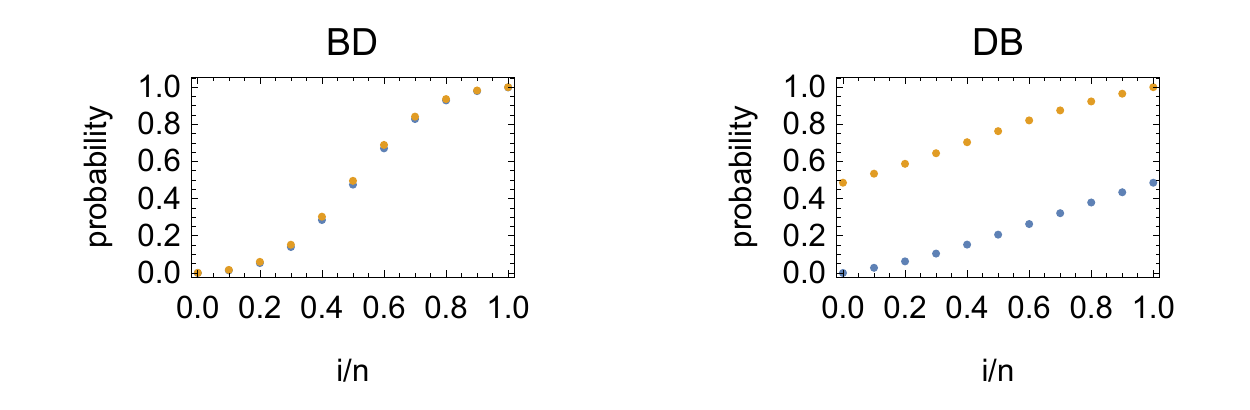}
		\caption{\label{figcoord} As in Fig. \ref{figindep}, the left and right panels refer respectively BD and DB updatings. Here we use the pay-off matrix (\ref{payoffcoord}) to define frequency-dependent fitnesses by (\ref{fi}) and (\ref{gi}). The number of leaves in the star graph is $n=10$. Again, as in Fig. \ref{figindep}, the difference between $P^0_i$ and $P^1_i$ is almost invisible in the BD plot. For the DB plot, the blue dots are the $P^0_i$ and the orange dots are the $P^1_i$ }
	\end{center}
	\end{figure}

We may now look again at (\ref{BDprobs}) and (\ref{DBprobs}) and understand why the order BD or DB of the lotteries is so important for the star graph. We let $n \rightarrow \infty$ and fix the fraction $x=i/(n+1)$ of A individuals in the population. In this limit, for fixed $x$, $i$ is of the order of $n$. Then in the BD case the probabilities $t_{i,\pm}$ involving drawing the center for reproduction are $O(1/n)$, i.e. small. On the contrary, the probabilities of drawing some leaf for reproduction are $O(1)$, and so are $t_{i,u}$ and $t_{i,d}$. The center in the BD case is very much influenced by the leaves. The reader may repeat a similar reasoning and see that in the DB case, on the contrary, the center influences very much the leaves. As a consequence, one should expect that whether the center is occupied by an A or a B should not influence very much the fixation probability of the A individuals in the BD case. On the contrary, we expect that in the DB case the occupation of the center by an A should increase substantially the fixation probability of the A individuals, and occupation of the center by a B should decrease substantially the A fixation probability. This strong difference between BD and DB is apparent in Figs. \ref{figindep} and \ref{figcoord}.

In order to solve the set of equations (\ref{eqp0i}) and (\ref{eqp1i}), we start by defining the differences
\[	d_i^0= P_i^0-P_{i-1}^0, \;\;\; d_i^1= P_i^1-P_{i-1}^1\]
and
\[d_i^{10}= P_i^1-P_i^0 \;.\]

We may then rewrite equations (\ref{eqp0i}) and (\ref{eqp1i}) respectively as
\begin{eqnarray}
d^0_i&=&\frac{1-\beta_i}{\beta_i}d^{10}_i\nonumber\\
d^1_i&=&\frac{1-\alpha_{i-1}}{\alpha_{i-1}}d^{10}_{i-1}\nonumber
\end{eqnarray}
and also, by the definitions of the differences, we get
\begin{equation*}
d^{10}_i=d^1_i-d^0_i+d^{10}_{i-1}\;.
\end{equation*}

We may consider the last three equations as a linear system for obtaining $d_i^0$, $d_i^1$ and $d_i^{10}$ all in terms of $d_{i-1}^{10}$. Solving it, we get
\begin{eqnarray}
d^1_i&=&\frac{1-\alpha_{i-1}}{\alpha_{i-1}}d^{10}_{i-1}\nonumber\\
d^{10}_{i}&=&\frac{\beta_i}{\alpha_{i-1}}d^{10}_{i-1}\label{soldiff}\\
d^0_i&=&\frac{1-\beta_i}{\alpha_{i-1}}d^{10}_{i-1}\nonumber
\end{eqnarray}

Solving the recursion given by the second of (\ref{soldiff}), we get
\begin{eqnarray}
d^{10}_i&=&\prod_{j=1}^{i}\left(\frac{\beta_j}{\alpha_{j-1}}\right)d_0^{10}\nonumber\\
&=&\prod_{j=1}^{i}\left(\frac{\beta_j}{\alpha_{j-1}}\right)P^1_0 \label{sold10i}\;,
\end{eqnarray}
because, by the first of (\ref{bc}), we have $d^{10}_0= P^1_0-P^0_0=P^1_0$. Observe here that this formula proves that for all $i$ we have $P^1_i>P^0_i$.

Substituting (\ref{sold10i}) in the first of (\ref{soldiff}), we have
\begin{eqnarray}
d^1_i&=&\frac{1-\alpha_{i-1}}{\alpha_{i-1}}\prod_{j=1}^{i-1}\left( \frac{\beta_j}{\alpha_{j-1}} \right)P^1_0\nonumber\\
&=& \frac{1-\alpha_{i-1}}{\alpha_0}\, \prod_{j=1}^{i-1} \frac{\beta_j}{\alpha_{j}}\,P^1_0\;.\label{sold1i}
\end{eqnarray}

An explicit formula for the $P^1_0$ can now be found, because, due to the second in (\ref{bc}), 
\begin{eqnarray}
1&=&P_n^1= d_n^1+d_{n-1}^1+ \dots +d_1^1+P^1_0\nonumber\\
&=& \frac{P_0^1}{\alpha_0}\left(1+ \sum_{j=1}^{n-1}(1-\alpha_j)\prod_{k=1}^j \frac{\beta_k}{\alpha_k}\right)\nonumber\;.
\end{eqnarray}
Solving this for $P_0^1$, we get
\begin{equation}\label{solP10}
P^1_0=\frac{\alpha_0}{1+\sum_{j=1}^{n-1}(1-\alpha_j)\prod_{k=1}^{j}\frac{\beta_k}{\alpha_k}}\;.
\end{equation}
The $P^1_i$, $i=1,2, \dots, n-1$, may be obtained recursively by $P^1_i=P^1_{i-1}+d^1_i$ and using (\ref{sold1i}) and (\ref{solP10}). The result is
\begin{equation} \label{solP1i}
P^1_i=\frac{1+\sum_{j=1}^{i-1}(1-\alpha_j)\prod_{k=1}^{j}\frac{\beta_k}{\alpha_k}}{1+\sum_{j=1}^{n-1}(1-\alpha_j)\prod_{k=1}^{j}\frac{\beta_k}{\alpha_k}}\;.
\end{equation}

Finally, we may calculate the $P^0_i$, $i=1,2, \dots, n$ as $P^1_i-d^{10}_i$. Using (\ref{solP1i}), (\ref{sold10i}) and (\ref{solP10}) we obtain
\begin{eqnarray}
P^0_i&=&\frac{1+\sum_{j=1}^{i-1}(1-\alpha_j)\prod_{k=1}^{j}\frac{\beta_k}{\alpha_k}- \beta_i \prod_{j=1}^{i-1} \frac{\beta_j}{\alpha_j}}{1+\sum_{j=1}^{n-1}(1-\alpha_j)\prod_{k=1}^{j}\frac{\beta_k}{\alpha_k}} \nonumber\\
&=& \frac{\sum_{j=1}^{i}(1-\beta_j) \prod_{k=1}^{j-1}\frac{\beta_k}{\alpha_k}} {1+\sum_{j=1}^{n-1}(1-\alpha_j)\prod_{k=1}^{j}\frac{\beta_k}{\alpha_k}} \;.
\label{solP0i}
\end{eqnarray}

Formulae (\ref{solP10}), (\ref{solP1i}) and (\ref{solP0i}) are the exact and explicit solution for the fixation probability of A individuals at any configuration in the star graph for frequency-dependent fitnesses and either BD or DB updating. If Ewens \cite{ewens} termed ``unwieldy" the analogous and simpler formula (\ref{exactpii}) for the unstructured population, then these formulae deserve much more a better comprehension. Before we do that in the next section, repeating some work similar to \cite{graphshapes} for the unstructured case, let us simply plot the results of (\ref{solP10}), (\ref{solP1i}) and (\ref{solP0i}) in two illustrative cases and both update rules.

In the first example, we take $f_i=r$, $g_i=1$, with $r>0$, so that $r_i=r$ in (\ref{defri}). This choice defines the transition probabilities in (\ref{BDprobs}) and (\ref{DBprobs}) to be substituted in (\ref{defalphabeta}). The interpretation of $r$ is that in the reproduction lottery the probability of a single chosen A individual being drawn is $r$ times the probability of a single chosen B be drawn. If $r>1$, A individuals are fitter, if $0<r<1$, Bs are fitter. The case $r=1$ is called \textit{neutral}. As $r$ is independent of $i$, we are in the simpler context of \textit{frequency-independent} fitness.

Fig. \ref{figindep} illustrates the behavior of the fixation probability of A individuals as a function of the number of A individuals in the leaves of the star graph, both for BD and DB. We take frequency-independent fitness with $r=1.2$.

For a general pay-off matrix $M$, in which types A and B are numbered respectively as 1 and 2, $m_{k \ell}$ is the pay-off of type $k$ when interacting with type $\ell$. The standard Evolutionary Game Theory \cite{nowakbook, graphshapes} fitnesses are 
\begin{eqnarray}
\label{fi}
f_i &=& m_{11} \, \frac{i-1}{N-1} \,+\, m_{12}\, \frac{N-i}{N-1} \\
\label{gi}
g_i &=&m_{21} \, \frac{i}{N-1} \,+\, m_{22}\, \frac{N-i-1}{N-1} \;.
\end{eqnarray}  
In the second example, illustrated at Fig. \ref{figcoord}, the pay-off matrix is 
\begin{equation}
\label{payoffcoord}
M=\left(
\begin{array}{cc}
10 & 5 \\
6 & 10 \\
\end{array}
\right) \;,
\end{equation}
i.e. a \textit{coordination} or \textit{stag hunt} game.

 \section{Asymptotics}\label{secAsymp}
 Taking the ratio between (\ref{fi}) and (\ref{gi}), it is easy to see that 
 \begin{equation}
 \label{rR}
 r_i=R(i/n)+O(1/n)\;,
 \end{equation}
 where 
 \begin{equation}
 \label{defR}
 R(x)=\frac{m_{11}x + m_{12} (1-x)} {m_{21}x+m_{22}(1-x)}
 \end{equation}
 is independent of $n$. 
 
 Let $x \in [0,1]$ be a fixed fraction of A individuals in the leaves of the star graph and $[n x]$ be the integer closest to $nx$. We define the asymptotic fixation probabilities as 
 \begin{equation*}
 	\pi^0 (x)=\lim\limits_{n \rightarrow \infty} P^0_{[nx]} \;\;\; \textrm{and} \;\;\; \pi^1 (x)=\lim\limits_{n \rightarrow \infty} P^1_{[nx]}\;.
 \end{equation*}
 
 In (\ref{solP10}), (\ref{solP1i}) and (\ref{solP0i}) the ratio $\alpha_k/\beta_k$ assumes a role similar to the relative fitness $r_k$ in the unstructured case (\ref{exactpii}). The strong qualitative differences between BD and DB seen in Figs. \ref{figindep} and \ref{figcoord} are justified by the fact that $\alpha_k/\beta_k$ is so different in the two cases when $n \rightarrow \infty$. We now separate the two cases.
 
 \subsection{DB case} 
Using  (\ref{defalphabeta}) and (\ref{DBprobs}), in the DB case we get
 \begin{equation}  \label{DBratio}
 	\frac{\alpha_i}{\beta_i}=1+ \ \frac{1}{n} \, \frac{r_i-1}{1+ \frac{i}{n}(r_i-1)}+O(\frac{1}{n^2})\;,
 \end{equation} 
 \begin{eqnarray} 
 	1-\alpha_i&=&\frac{1}{n} \ \frac{1}{1+\frac{i}{n}(r_i-1)}+O(\frac{1}{n^2}) \label{1-aDB}\\ 
 	1-\beta_i&=&\frac{1}{n} \ \frac{r_i}{1+\frac{i}{n}(r_i-1)}+O(\frac{1}{n^2})\label{1-bDB}\;.
 \end{eqnarray}

We proceed by writing the product of the $\alpha_k/\beta_k$ as an exponential of a sum. Because, by (\ref{DBratio}), $\log(\alpha_k/\beta_k)$  is $O(1/n)$, the sum conveniently converges to an integral:
\begin{eqnarray*}
	\prod_{k=1}^{[nx]}\frac{\beta_k}{\alpha_k}&=&\exp \left[- \sum_{k=1}^{[nx]}\log \frac{\alpha_k}{\beta_k}\right] \nonumber\\
	&=&\exp \left[-\sum_{k=1}^{[nx]}\frac{1}{n}\left( \frac{r_k-1}{1+\frac{k}{n}(r_k-1)}+O(\frac{1}{n}) \right)\right]\\
	&\stackrel{n \rightarrow \infty}{\longrightarrow}&\exp\left[-\displaystyle\int_{0}^{x}\frac{R(z)-1}{1+z(R(z)-1)}  dz\right]\;.
\end{eqnarray*}
Observe now that the convenient $1/n$ leading behavior is present also in (\ref{1-aDB}) and (\ref{1-bDB}). So the sums in (\ref{solP1i}) and (\ref{solP0i}) converge to integrals, too. 

We may then define functions which are the limits of the sums appearing in the numerators and denominators of (\ref{solP1i}) and (\ref{solP0i}):
\begin{equation*}
	\varTheta(x)\equiv \int_{0}^{x}\frac{R(y)}{1+y \ (R(y)-1)} \ e^{-\int_{0}^{y}\frac{R(z)-1}{1+z(R(z)-1)} \ dz}\ dy
\end{equation*}
and
\begin{equation*}
	\varXi(x)\equiv\int_{0}^{x}\frac{1}{1+y \ (R(y)-1)} \ e^{-\int_{0}^{y}\frac{R(z)-1}{1+z(R(z)-1)} \ dz}\ dy\;.
\end{equation*}
In terms of these functions, we prove that in the DB case
\begin{equation*}
	\pi^0(x)\,=\, \frac{\varTheta(x)}{1+\varXi(1)}\;\;\; \textrm{and}\;\;\;	\pi^1(x)\,=\,\frac{1+\varXi(x)}{1+\varXi(1)}\;.
\end{equation*}
 
\subsection{BD case} 
The formulae analogous to (\ref{DBratio}-\ref{1-bDB}) in the BD case are 
 \begin{equation}\label{BDratio}
 \frac{\alpha_i}{\beta_i}=r_i\, r_{i+1}\left[1+\frac{1}{n}\, \frac{1-r_i r_{i+1}}{r_i}+O(\frac{1}{n^2})\right]\;,
 \end{equation}
 \begin{eqnarray}
 1-\alpha_i&=& 1-\frac{r_{i+1}}{n}+O(\frac{1}{n^2})\label{1-aBD} \\ 1-\beta_i&=&1-\frac{1}{n\,r_i}+O(\frac{1}{n^2})\label{1-bBD}\;.
 \end{eqnarray}
 
Lieberman et al. \cite{lieberman} had already noticed -- for the BD updating and frequency-independent fitness -- that the fixation probability of a single A individual in a star graph is asymptotically equal to the fixation probability in an unstructured population with relative fitness $r$ replaced by $r^2$. This replacement is responsible for the fact that the star is an amplifier of selection. At first sight, if we take $r_i=r$, neglect the corrections tending to 0 when $n \rightarrow \infty$ in (\ref{BDratio}-\ref{1-bBD}) and use the exact formula (\ref{solP1i}), we see such a result.

But the above argument is not strictly true, because Chalub \cite{chalubstar} noticed that the asymptotic expression of Lieberman et al. has to be corrected. The source for this correction is that the $O(1/n)$ contribution in (\ref{BDratio}) cannot be simply neglected. The analysis of the BD case is more involved, but similar to the asymptotics for the unstructured population case, thoroughly explained in \cite{graphshapes}. Part of the same analysis had been done before by Antal and Scheuring \cite{AntalScheuring}, but \cite{graphshapes} is more complete and also corrects some mistakes. In the following we will try to give a reasonable account of this analysis, referencing the reader to the cited papers for the technical details.

From a mathematical point of view, the main difference between BD and DB is that, contrary to the DB case,  $\log(\alpha_k/\beta_k)$ is $O(1)$, not $O(1/n)$. This is what makes the BD case similar to the unstructured case. 

For $x \in [0,1]$ and $R$ given by (\ref{defR}), define, as in \cite{graphshapes}, the \textit{fitness potential}
\begin{equation}
\label{defL}
L(x) \,\equiv  \, - \, \int_{0}^{x} \log R(t) \,dt \;.
\end{equation}
A similar definition was given by Chalub and Souza \cite{chalubsouza1} and further explored by the same authors in \cite{chalubsouza2}.
It can be seen that $L$ always has a single maximum point $x^* \in [0,1]$. The location of $x^*$ in the interior of the interval, or in one of its boundary points depends on the \textit{invasion scenario} \cite{taylor, graphshapes}, i.e. on the possible signs of $R(0)-1$ and $R(1)-1$.

In order to deal with the correction by Chalub, related to the $O(1/n)$ terms in (\ref{BDratio}), we define also
\begin{equation}
\label{defC}
C(x) \,\equiv  \, \int_{0}^{x} \left(\frac{R(t)^2-1}{R(t)}- \frac{\det M}{m_{12}m_{22}}\right) \,dt \;.
\end{equation}

Using a reasoning similar to the DB case, we write
\begin{eqnarray}
	\prod_{k=1}^{[nx]}\frac{\beta_k}{\alpha_k}&=&\exp \left[-\sum_{k=1}^{[nx]}\log \frac{\alpha_k}{\beta_k}\right] \nonumber\\
	&=&\exp\left[-n \sum_{k=1}^{[n x]}\frac{1}{n} (\log r_k+\log r_{k+1}) \right]\, \exp \left[-\sum_{k=1}^{[nx]}\frac{1}{n}\left( \frac{1-r_k r_{k+1}}{r_k}+O(\frac{1}{n}) \right)\right]\nonumber\\
	&=& \exp\left[-2n \sum_{k=1}^{[n x]}\frac{1}{n} \log r_k \right]\, \exp \left[-\sum_{k=1}^{[nx]}\frac{1}{n}\left( \frac{1-r_k r_{k+1}}{r_k}+ \frac{\det M}{m_{12}{m_{22}}}+O(\frac{1}{n}) \right)\right] \nonumber \\
	&\stackrel{n \rightarrow \infty}{\sim}& e^{C(x)+E(x)}\, e^{2nL(x)} \label{asympprod} \;,
\end{eqnarray}
where the $E(x)$ term will be explained below.

In the above asymptotic formula, the main term is $e^{2n L(x)}$. It appears both in \cite{graphshapes, AntalScheuring} and also, as $r^{-2n}$, for the frequency-independent fitnesses case in \cite{lieberman}. The $e^{C(x)}$ generalizes the correction by Chalub \cite{chalubstar}, and the $e^{E(x)}$ is of the type called ``continuation error" in \cite{graphshapes}, because it appears when the sum $\sum_{k=1}^{[n x]}\frac{1}{n} \log r_k$ is replaced by the integral $-L(x)$.

The first thing to be explained in the BD case is why the difference $P^1_i-P^0_i = d^{10}_i$ is so small when $n$ is large. This can be seen by substituting (\ref{solP10}) in (\ref{sold10i}) and obtaining
\begin{equation}\label{altd10i}
d^{10}_i = \frac{\beta_i \, \prod_{j=1}^{i-1} \frac{\beta_j}{\alpha_j}} {1+\sum_{j=1}^{n-1}(1-\alpha_j)\prod_{k=1}^{j}\frac{\beta_k}{\alpha_k}}\;,
\end{equation}
which holds also in the DB case.

Using the asymptotic expression (\ref{asympprod}), if $n$ is large, then the sum in the denominator of (\ref{altd10i}) is dominated by the term such that $L(j/n)$ is maximum, which tends to $x^*$ when $n \rightarrow \infty$. Using (\ref{asympprod}) also in the numerator of (\ref{altd10i}), we get for large $n$ and fixed $x \in [0,1]$,
\[d^{10}_{[n x]} \approx \frac{1}{n} \, c_n(x) \, e^{2n (L(x)-L(x^*))} \;,\]
where the $1/n$ factor comes from the $\beta_i=1/(n r_i+1)$ in the numerator and $c_n(x)$ may be either $O(1)$ if $x^*$ is a boundary point, or $O(n^{-1/2})$ if $x^*$ is an interior point. We omit here some technicalities which can be found in \cite{graphshapes}. In any case, if $x \neq x^*$, $d^{10}_{[n x]}$ is exponentially small in $n$, and, even if $x=x^*$, it tends to 0 as $n \rightarrow \infty$, although more slowly. In the example plotted in the left panel of Fig. \ref{figcoord} we have $x^*=5/9$, which locates correctly the region in which the difference $P^1_i-P^0_i$ is more visible.

Having accepted that the difference between $P^1_i$ and $P^0_i$ is small, we may then use the expression (\ref{solP1i}) for the former to approximate both. We use it, because we can now approximate $1-\alpha_j$ by 1, see (\ref{1-aBD}), and (\ref{solP1i}) takes the same form as (\ref{exactpii}) with $r_k$ replaced by $\alpha_k/ \beta_k$. The possible graph shapes and asymptotic behavior for (\ref{exactpii}) were studied in \cite{graphshapes} and hold here as approximations for the $P^1_i$ and $P^0_i$ in the BD case. In particular, for fixed $x \in(0,1)$, $P^1_{[n  x]}$ and $P^0_{[n x]}$ both tend to 0 when $n \rightarrow \infty$ if $x<x^*$ and to 1 if $x>x^*$.

\section{Conclusions}\label{secconc}
The discovery of an exact solution for a simplified problem, as the Onsager solution \cite{onsager} for the two-dimensional Ising model in Statistical Mechanics,  is an important result. In fact, one can proceed to more realistic models and develop approximation techniques based on the increased understanding gained by the exact solution. In this paper we have generalized the exact solution found by Broom and Rycht\'{a}\v{r} \cite{broomrychtar} and derived by another method by Monk et al. \cite{monk}. Both papers provide explicit formulae only when the fitnesses of A and B individuals are frequency-independent and the updating rule is BD.

Our formulae (\ref{solP10}), (\ref{solP1i}) and (\ref{solP0i}) are rather complicated, but we can understand them very well in the important asymptotic limit when the number of leaves $n$ in the star graph tends to $\infty$. We see rather important differences between the DB and BD cases. In the DB case we obtain asymptotic formulae in terms of integrals. The BD case is harder, but it can be understood by techniques introduced elsewhere \cite{AntalScheuring, graphshapes}.

We hope that the results of this paper may encourage other researchers to understand more thoroughly the fixation probabilities of the Moran process in more general graphs.

\section*{Acknowledgments}
	This study was financed in part by the Coordena\c{c}\~ao de
	Aperfei\c{c}oamento de Pessoal de N\'ivel Superior - Brasil (CAPES) -
	Finance Code 001.


\bibliographystyle{plain}
\bibliography{bibexactstar}

\end{document}